\documentclass[useAMS,usenatbib,a4paper,preprint]{mn2e}
\usepackage{graphicx}
\usepackage{amsmath}
\usepackage{txfonts}
\usepackage{mathrsfs}
\usepackage{color}

\def\aap{A\&A}
\def\apjl{ApJ}

\def\apjs{ApJS}

\newcommand{\be}{\begin{equation}}
\newcommand{\ee}{\end{equation}}
\newcommand{\bea}{\begin{eqnarray}}
\newcommand{\eea}{\end{eqnarray}}

\title[Cosmic Distance Duality Relation]{A comparison of the $R_{\rm h}=ct$ and
$\Lambda$CDM cosmologies using the Cosmic Distance Duality Relation} 
  \author[Fulvio Melia]
    {Fulvio Melia\thanks{John Woodruff Simpson Fellow. Email: fmelia@email.arizona.edu} \\
     Department of Physics, The Applied Math Program, and Department of Astronomy,
     The University of Arizona, AZ 85721, USA}

\begin{document}

\date{}

\pagerange{\pageref{firstpage}--\pageref{lastpage}} \pubyear{2016}

\maketitle

\label{firstpage}

\begin{abstract}
The cosmic distance duality (CDD) relation (based on the Etherington 
reciprocity theorem) plays a crucial role in a wide assortment of cosmological 
measurements. Attempts at confirming it observationally have met with mixed results, 
though the general consensus appears to be that the data do support its existence in 
nature. A common limitation with past approaches has been their reliance on a specific 
cosmological model, or on measurements of the luminosity distance to Type Ia SNe, 
which introduces a dependence on the presumed cosmology in spite of 
beliefs to the contrary. Confirming that the CDD is actually realized in nature is 
crucial because its violation would require exotic new physics. In this paper, 
we study the CDD using the observed angular size of compact quasar 
cores and a Gaussian Process reconstruction of the HII galaxy Hubble diagram---without 
pre-assuming any particular background cosmology. In so doing, we confirm at a very 
high level of confidence that the angular-diameter and luminosity distances
do indeed satisfy the CDD. We then demonstrate the potential power of 
this result by utilizing it in a comparative test of two competing cosmological 
models---the $R_{\rm h}=ct$ universe and $\Lambda$CDM---and show that 
$R_{\rm h}=ct$ is favoured by the CDD data with a likelihood $\sim 82.3\%$ 
compared with $\sim 17.7\%$ for the standard model.
\end{abstract}


\section{Introduction}
The so-called cosmic distance duality (CDD) relation, based on the
reciprocity theorem first derived by Etherington (1933) (and revived by Ellis 1971),
holds true as long as (i) the cosmic spacetime is based on Riemannian geometry, (ii) 
photons propagate along null geodesics, and (iii) the photon number is conserved. 
Mathematically, the CDD may be expressed in the form $\eta(z)=1$, where
\begin{equation}
\eta(z) = (1+z)^2{d_A(z)\over d_L(z)}\;,
\end{equation}
in terms of the angular-diameter distance $d_A(z)$ and luminosity distance $d_L(z)$.
There are many reasons why the CDD could in principle be violated 
in nature, at least one of which was invoked to test non-standard cosmologies
by Bassett \& Kunz (2004a). The violation could arise if the spacetime is not described by 
a metric theory of gravity, which many believe is unlikely (see, e.g., Adler 1971; Bassett 
\& Kunz 2004b), or perhaps because photons arriving from presumed standard candles, such 
as Type Ia SNe, are altered by absorption or scattering along the line of sight (see, 
e.g., Sikivie 1983; Bassett \& Kunz 2004a; Raffelt 1999; Chen 1995; Deffayet \& Uzan 2000; Khoury
\& Weltman 2004; Burrage 2008; Liao et al. 2015).

Following early papers on this topic by Bassett \& Kunz (2004a, 2004b),
many have attempted to validate the CDD using an array of observational
data, typically using angular-diameter distances extracted from galaxy clusters
and luminosity distances from Type Ia SNe (for a non-exhaustive list, see
Uzan et al. 2004; Bernardis et al. 2006; Holanda et al. 2010, 2012; Khedekar 
\& Chakraborti 2011; Li et al. 2011; Nair et al. 2011; Meng et al. 2012; 
Ellis et al. 2013; Liao et al. 2016; Ma \& Corasaniti 2016; Hu \& Wang 2018).
The issue of some cosmology dependence in these measurements 
is not a trivial one, however. There is always the possibility that 
the assumption of a specific cosmological model biases the distances, particularly 
if the model is incomplete (or even wrong). Under such circumstances, tests of the 
CDD may not produce compelling, incontrovertible outcomes. For this reason, some 
(or all) violations of the CDD claimed by previous studies may simply be due to 
unaccounted for influences of the assumed cosmology (see, e.g., Uzan et al. 2004;
Holanda et al. 2010; Li et al. 2011). 

Several earlier workers have claimed to be testing the CDD in a 
manner that was free of any possible dependence on the cosmology, 
believing that the use of luminosity distances extracted from Type Ia SNe
does not rely on any particular expansion scenario (see, e.g., 
Holanda et al. 2010, 2012; Meng et al. 2012; Liao et al. 2016). But the reality 
is that in order to `turn' Type Ia SNe into standard candles, one must 
simultaneously optimize the unknown parameters describing their lightcurve 
along with the parameters of the presumed cosmological model 
(see, e.g., Amanullah et al. 2010). These three or four (depending on the
application) so-called `nuisance' parameters do not exist in isolation and
cannot be identified properly without training the lightcurve fitter in
the context of a particular cosmology to relate the lightcurves of SNe at different
redshifts (for a discussion relevant to this see, e.g., Kim 2011; Yang et al. 
2013; Wei et al. 2015c). It is now very well understood that Type Ia SN data 
reduced with $\Lambda$CDM as the background cosmology cannot be used for other 
models; one must of necessity re-optimize all of the parameters, including 
those of the lightcurve fitter, separately for each different expansion scenario. 

In this paper, we attempt to test the CDD with as little dependence 
on cosmological models as possible, using a new approach for measuring the 
angular-diameter and luminosity distances. As we shall see, our method
confirms the CDD at a much higher level of confidence than was achieved
before. We use the measured angular size of compact quasar cores 
as an indicator of the angular-diameter distance, and a Gaussian Process
reconstruction of the HII galaxies Hubble diagram to determine the luminosity
distance. In neither case is it necessary to assume a cosmological model
beforehand. As we shall see, the CDD measured in this fashion may 
then be used to test individual cosmologies. We introduce the data 
in \S~2, and analyze the distance duality relation in \S~3, confirming 
that it is satisfied at a very high level of confidence. As an example of 
the usefulness of this result, we then apply it to one-on-one model
selection in \S~4. We present our conclusion in \S~5.

\section{Data}
In this section, we describe two sets of data suitable for testing
the CDD (derived from Etherington's reciprocity relation), one 
based on the angular size of compact quasar cores, from which we obtain 
$d_A(z)$, the other a Hubble diagram constructed with HII galaxies (HIIGx) 
and Giant extragalactic HII regions (GEHR) as standard candles, from which
we infer $d_L(z)$. As we shall see, this test allows us to determine
the redshift dependence of $\eta$ without the need to pre-assume any 
cosmological model. 

\subsection{Angular size of compact quasar cores}
Recent improvements in our understanding of compact quasar 
cores make it possible for us to identify a luminosity and spectral-index 
limited sample of central, opaque regions in these sources and use them as 
reliable measuring rods. These cores probe the geometry of the Universe over 
a much bigger fraction of its age (corresponding to $0\lesssim z\lesssim 3$)
than even Type Ia SNe can currently achieve. We have recently used these data 
to measure the redshift $z_{\rm max}$ at which the angular-diameter distance 
$d_A(z)$ reaches its maximum value (Melia 2018a; Melia \& Yennapureddy 2018), 
finding that $z_{\rm max}=1.70 \pm0.20$---a unique new measure of the cosmic 
expansion (see also Cao et al. 2017). The location of this turning point is 
a strong function of the underlying cosmology and may be used for model 
selection, whose results thus far have strongly favoured the $R_{\rm h}=ct$ 
universe (Melia 2007; 2013b, 2016a, 2017a; Melia \& Abdelqader 2009; Melia 
\& Shevchuk 2012), followed by {\it Planck} $\Lambda$CDM (Planck Collaboration 
2016). Several other models, including (and especially) Milne (see, 
e.g., Vishwakarma 2013; Chashchina \& Silagadze 2015), Einstein-de 
Sitter (see, e.g., Vauclair et al. 2003; Blanchard 2006) and Static 
tired light (see, e.g., La Violette 2012) have been strongly rejected.

Given the history with the use of quasars and radio galaxies
to probe the cosmological expansion, one should view the use of compact
quasar cores to measure angular-diameter distance with some caution.
In this paper, we demonstrate how the CDD may be tested
using data such as these without the need to adopt any particular
cosmological model. Nevertheless, this analysis is strongly dependent
on astrophysical processes, such as the radio emission from quasar
cores, that may not be completely free of unknown systematics. We
must therefore emphasize that when we refer to the method employed
here as being ``model-independent," this designation refers solely to 
the cosmological background, not necessarily the physics of self-absorbed
synchrotron emission in quasars and the HII line emission we shall
discuss shortly.

Based on what we now know about these jet sources, we understand that 
their base emission is dominated by self-absorbed synchrotron radiation
(Blandford \& K\"onigl 1979; Melia \& K\"onigl 1989; Melia et al. 1992;
Nayakshin \& Melia 1998; Liu \& Melia 2001; B\'elanger et al. 2004; 
Chan et al. 2009; Crocker et al. 2011; Trap et al. 2011), creating 
optically-thick structures with angular sizes in the milliarcsecond (mas) 
range. Their corresponding physical length ($\sim$$10$ parsecs) is much 
smaller than the large-scale environment of the host galaxies. In addition, 
their morphology and kinematics are regulated by just a handful of parameters 
linked to the central engine (e.g., the mass and spin), and typically last 
only tens of years (Gurvits, Kellermann \& Frey 1999). One may
therefore reasonably assume that the structure and size of compact
quasar cores are independent of any long-term evolutionary effects
in the hosts themselves (Kellermann 1993; Jackson 2004, 2008).
But though the idea of using compact radio sources for the optimization 
of cosmological parameters was proposed almost 3 decades ago
(see, e.g., Kellermann 1993), a persistent complication has
been that they comprise a mixture of quasars, OVVs and BL Lacs,
among several others, making it difficult to disentangle systematic
differences among them from real cosmological variations.

This limitation notwithstanding, several significant 
improvements in selecting an appropriate sub-sample of these sources 
have produced a catalog suitable for use as standard rulers. The first was
a constraint on their spectral index $\alpha$ (Gurvits, Kellermann
\ Frey 1999), which led to the reduced sample assembled by Jackson
\& Jannetta (2006), which itself was extracted from an old 2.29 GHz
VLBI survey of Preston et al. (1985), with additions by Gurvits
(1994) (see also Jackson \& Dodgson 1997). 
More recently, Cao et al. (2017) followed the lead established
earlier by Gurvits, Kellermann \& Frey (1999) and Vishwakarma (2001)
in analyzing a possible mitigation of the scatter in core size by not 
only restricting their spectral index, but also their luminosity $L$.
These authors showed that adopting the parametrization 
${\ell}_{\rm core}={\ell}_0\,L^\gamma(1+z)^n$, for the core size
${\ell}_{\rm core}$ in terms of a scaling constant ${\ell}_0$,
one could attain a remarkably uniform sample with 
$\gamma\approx 10^{-4}$ and $|n|\approx 10^{-3}$, by simply
choosing only intermediate-luminosity radio quasars in the range
$10^{27}\,{\rm W/Hz}<L<10^{28}\,{\rm W/Hz}$ with spectral index
$-0.38<\alpha<0.18$.

Figure~1 shows the data used in this paper, following the selection
procedure described above. These are drawn from the original
613 sources of Jackson \& Jannetta (2006)\footnote{The full
sample is also available at  http://nrl.northumbria.ac.uk/13109/}. 
Using the {\it Planck} 
optimized parameters (Planck Collaboration 2016) to estimate
$d_L(z)$, the flux density at $2.29$ GHz yields the luminosity 
$L$, from which we extract the sub-sample with intermediate 
luminosities. Note that this step merely estimates $L$ for the
purpose of source selection. The {\it Planck} parameters 
are not used in any other way, so the results do not
depend on the parametrization in $\Lambda$CDM. A subsequent
restriction of the spectral index $\alpha$ generates
the final catalog of 140 sources used in this paper.
These sources are then binned into groups of 7, and the 
median value is chosen in each bin to represent the
core angular size $\theta_{\rm core}(z)$ (Santos
\& Lima 2008), with associated $1\sigma$ errors estimated 
assuming Gaussian scatter within each bin. This step 
partially minimizes the scatter that would otherwise appear
with individual data points, but note that it does not at
all reduce the measurement uncertainty, here represented by 
the standard deviation $\sigma$. In other words, the 
remaining scatter in the individual data points is reflected
in the size of the error bars associated with the 20 data 
points in this figure.

\begin{figure}
\vskip 1cm
\centerline{
\includegraphics[angle=0,scale=0.7]{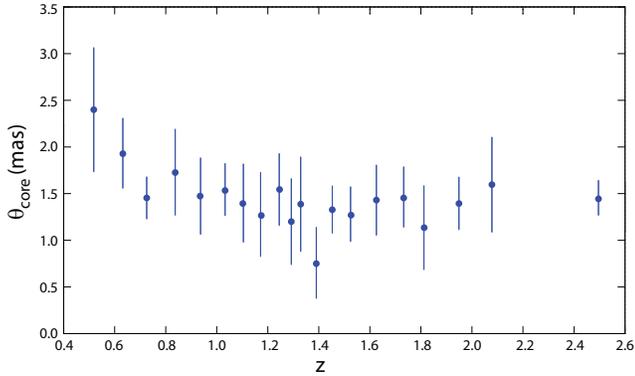}}
\vskip 0.2in
\caption{Angular size of 140 compact quasar cores divided into
bins of 7, as a function of redshift. Each datum represents
the median value in its bin, and its standard deviation is 
estimated assuming Gaussian variation within each bin. (Adapted 
from Melia 2018a)}
\end{figure}

Writing the angular size of a compact quasar core as
\begin{equation}
\theta_{\rm core}(z) = {\ell_{\rm core}\over d_A(z)}\;,
\end{equation}
where $d_A(z)$ is the aforementioned angular-diameter distance,
we see that, as long as the physical core size ${\ell}_{\rm core}$ 
is approximately constant in the reduced quasar sample, the 
measured $\theta_{\rm core}(z)$ is a valid (inverse) representation
of the redshift-dependent angular-diameter distance, independent
of any cosmological model. We shall describe shortly how to
use the CDD in a way that avoids the need to know the actual value of 
${\ell}_{\rm core}$.

\subsection{HII Galaxy Hubble diagram}
HII galaxies (HIIGx) and Giant extragalactic HII regions (GEHR) have
similar optical spectra and massive star formation (Melnick et al. 1987)
and their hydrogen gas, ionized by massive star clusters, emits prominent 
Balmer lines in H$\alpha$ and H$\beta$ (Searle \& Sargent 1972;
Bergeron 1977; Terlevich \& Melnick 1981; Kunth \& {\"O}stlin 2000).
The luminosity $L({\rm H}\beta$) in H$\beta$ in these structures is strongly 
correlated with the velocity dispersion $\sigma_v$ of the ionized gas 
(Terlevich \& Melnick 1981), apparently because both the intensity of ionizing 
radiation and $\sigma_v$ increase with the starburst mass (Melnick et al.
2000; Siegel et al. 2005). Not surprisingly, the relatively small dispersion 
in the relationship between $L({\rm H}\beta$) and $\sigma_v$ allows these 
galaxies and HII regions to be used as standard candles (Melnick et al. 1987, 1988;
Fuentes et al. 2000; Bosch et al. 2002; Telles 2003; Siegel et al. 2005;
Bordalo \& Telles 2011; Plionis et al. 2011; Mania \& Ratra 2012; Chavez et al.
2012, 2014; Terlevich et al. 2015; Wei et al. 2016).

All these previous applications of the $L({\rm H}\beta$) - $\sigma_v$ correlation,
however, were based on parametric fits linked to specific cosmological models.
Were we to repeat this procedure here for the calculation of the luminosity
distance, $d_L(z)$, our results would similarly depend on a particular
expansion scenario. Instead, we follow the approach introduced by Yennapureddy 
\& Melia (2017), in which the function representing the HIIGx and GEHR data is 
reconstructed using Gaussian Processes (GP)---without the 
pre-assumption of any particular cosmology (Seikel et al. 2012). With 
this novel statistical method, one may reconstruct the function that 
best fits the data without assuming any parametric form at all. 

\begin{figure}
\centerline{
\includegraphics[angle=0,scale=0.5]{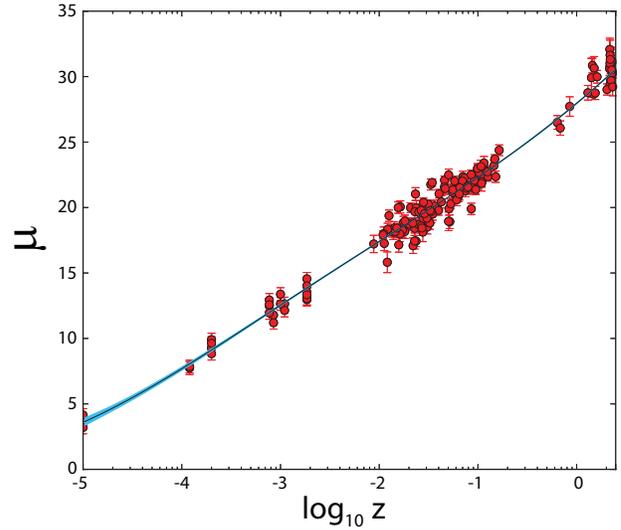}}
\vskip 0.2in
\caption{Distance modulus of the 156 currently available HII-region/Galaxy 
measurements, shown as red circles with $1\sigma$ error bars. The GP
reconstructed distance modulus $\mu^{\rm obs}(z)$ is shown as a thin black curve (see
text), and the blue swath represents the $1\sigma$ confidence region of
the reconstruction. (Adapted from Yennapureddy \& Melia 2017)}
\end{figure}

For this paper, we use the 25 high-$z$ HII galaxies, 107 local HII galaxies, 
and 24 giant extra galactic HII regions (compromising 156 sources in all)
from Terlevich et al. (2015). Their luminosity versus velocity dispersion
correlation may be written (Chavez et al. 2012, 2014; Terlevich et al. 2015)
\begin{equation}
\log L({\rm H}\beta)=\alpha \log \sigma_v ({\rm H}\beta) +\kappa\;,
\end{equation}
where $\alpha$ and $\kappa$ are constants. When fitting these data using a
particular cosmology, one must optimize these parameters simultaneously with
those of the model, but previous work has shown that they are quite insensitive
to the background cosmology (Wei et al. 2016). Combining $\kappa$ and $H_0$ 
together according to 
\begin{equation}
\delta =-2.5\kappa-5\log H_0 +125.2\;,
\end{equation}
this earlier analysis demonstrated that $\alpha$ and $\delta$ deviate from one 
model to the next by only a tiny fraction of their standard deviation. For example,
one gets $\alpha=4.86^{+0.08}_{-0.07}$ and $\delta=32.38^{+0.29}_{-0.29}$ 
when using the $R_{\rm h}=ct$ cosmology, compared with $\alpha=4.89^{+0.09}_{-0.09}$
and $\delta=32.49^{+0.35}_{-0.35}$ for $\Lambda$CDM. As noted, such small 
differences are well within the measurement error so, following our goal
of reconstructing the correlation function independently of any model, we 
shall simply adopt the average values reported by Wei et al. (2016) for 
these `nuisance' parameters, which are $\alpha=4.87^{+0.11}_{-0.08}$ and 
$\delta=32.42^{+0.42}_{-0.33}$. 

The distance modulus for an HII galaxy is then given as 
\begin{equation}
\mu^{\rm obs}=-\delta +2.5 \big[\alpha \log \sigma_v({\rm H}\beta)-\log F({\rm H}\beta)\big]\;,
\end{equation}
and since 
\begin{equation}
\mu^{\rm obs}(z)=5\log\bigg[\frac{d^{\,\rm obs}_L(z)}{\rm Mpc}\bigg]+25\;,
\end{equation}
we may write 
\begin{equation}
d^{\,\rm obs}_L(z) = {\rm const.}\;10^{\,\mu^{\rm obs}(z)/5}\;.
\end{equation}
The distance moduli for the 156 sources used in this study are plotted in
fig.~2, along with the GP reconstruction of the function representing them. 

A full description of the GP method, as applied to sources such as the
HIIGx and GEHR catalogs, appears in Yennapureddy \& Melia (2017, 2018a), 
based on the pioneering work of Seikel et al. (2012), and we refer
the reader to these publications for all the details. An important 
feature of the GP approach, of particular relevance to the analysis in this 
paper, concerns the estimation of the $1\sigma$ confidence region associated 
with the reconstructed $\mu^{\rm obs}(z)$ function, which is shown as a blue swath in
fig.~2. The width of this region depends on both the actual errors of 
individual data points and on the strength of the correlation function
used in the reconstruction (see, e.g., Seikel et al. 2012). The dispersion
at any redshift is less than the measured standard deviation at that point
when there is a large correlation in the reconstruction which, as it turns out,
is the situation we have here (Yennapureddy \& Melia 2017). The GP estimated 
$1\sigma$ confidence region is therefore smaller than the errors in the 
original data. This feature is one of the main benefits of using the GP 
approach to reconstruct the $L({\rm H}\beta$) - $\sigma_v$ correlation for 
this work. 

There is, however, an important caveat to keep in mind
with the use of these HII galaxy data and their GP reconstruction, having
to do with possible unknown systematics with the HII galaxy probe. The
$L(\mathrm{H}\beta) \text{-} \sigma$ correlation is still not completely understood.
There exist uncertainties in the size of the starburst, its age, the oxygen abundance 
in HII galaxies and also an internal extinction correction (Ch\'avez et al. 2016).
The scatter found with the use of Equation~(3) points to a possible dependence on
a second parameter. In their attempt to mitigate these uncertainties, Ch\'avez et al.
(2014) found that the size of the star-forming region can serve as this second
parameter. We should also keep in mind that the $L(\mathrm{H}\beta) \text{-} \sigma$
relation we are using ignores any rotating support for the system (as opposed to purely 
kinematic support). Ch\'avez et al. (2014, 2016) have proposed using an upper limit 
to the velocity dispersion in order to minimize this possible effect, but then the
the catalogue of suitable sources would be greatly reduced. In addition, there is
no guarantee that this systematic effect would be completely eliminated. We
therefore caution that, although we are not pre-assuming any particular
cosmological model for this analysis, the results we present in this paper are 
nonetheless subject to these possible weaknesses in our derivation of the distance 
modulus. The hope is that future improvements in our understanding of these systems 
will render the HII Hubble diagram an even more robust probe of the integrated 
distance measure than it is now.

\section{The Cosmic Distance Duality Relation}
We now have at our disposal 20 measurements of $d_A(z)$ based on the
angular size of compact quasar cores, ranging in redshift from $z=0$ to 
$z\sim 2.6$, and a GP reconstruction of the luminosity distance $d^{\,\rm obs}_L(z)$ 
from the HIIGx/GEHR Hubble diagram. Using our definition of $\eta(z)$
in the CDD (Eq.~1), we may assemble these data and write
\begin{equation}
\eta^{\rm obs}(z) = {\rm const.}\;\theta_{\rm core}^{-1}(z)\,10^{-\mu^{\rm obs}(z)/5}(1+z)^2\;.
\end{equation}
The constant in this expression includes the unknown physical scale 
${\ell}_{\rm core}$. Until this quantity is measured reliably, 
we are restricted in how well we can determine the absolute scaling 
$\eta^{\rm obs}(0)$. Nonetheless, we may use these data to measure 
the redshift dependence of the CDD without reference to any specific 
cosmological model. For this purpose, we write
\begin{equation}
\eta^{\rm obs}(z)=a+bz\;,
\end{equation}
and optimize the value of the constant $b$ using $\chi^2$ minimization
with the data shown in fig.~3. Given that we wish to avoid
having to use an actual measurement of ${\ell}_{\rm core}$,
we normalize the proportionality constant in Equation~(8) to yield $a=1$. 
The 20 data points plotted in fig.~3 correspond to this value of $a$. The 
associated errors are estimated using conventional error propagation,
\begin{equation}
\sigma_\eta=\eta^{\rm obs}\left[\left({\sigma_{d^{\,\rm obs}_A}\over d^{\,\rm obs}_A}\right)^2+
\left({\sigma_{d^{\,\rm obs}_L}\over d^{\,\rm obs}_L}\right)^2\right]^{1/2}\;,
\end{equation}
where $\sigma_{d^{\,\rm obs}_A}$ and $\sigma_{d^{\,\rm obs}_L}$ are the errors in 
$d^{\,\rm obs}_A$ and 
$d^{\,\rm obs}_L$, respectively, and we have assumed no covariance between these 
two measures of distance, given that they are based on two entirely different 
sets of data and analyses.

\begin{figure}
\centerline{
\includegraphics[angle=0,scale=0.75]{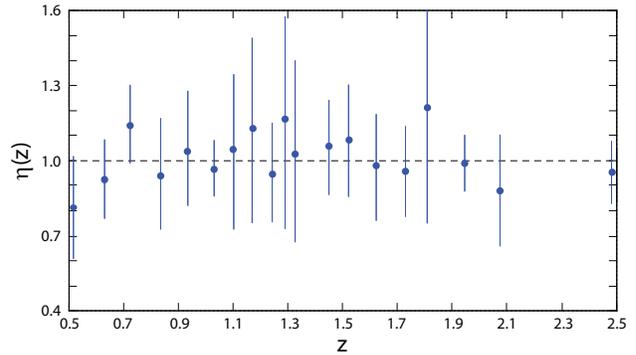}}
\vskip 0.2in
\caption{Values of $\eta$ estimated for the 20 measured quasar core
angular sizes shown in fig.~1 and the GP reconstructed Hubble diagram
in fig.~2. The horizontal dashed line (i.e., $\eta=1$) shows the value 
of $\eta$ for a perfectly satisfied CDD. These data
confirm the CDD at a very high level of confidence.}
\end{figure}

As summarized in fig.~4, our $\chi^2$ minimization yields the following result
for the fit using Equation~(9):
\begin{eqnarray}
\hskip1.0in a&=&1.00\pm0.05\nonumber\\
\hskip1.0in b&=&-0.01\pm0.03\;.
\end{eqnarray}
In other words, the data shown in fig.~3 are entirely consistent with no 
redshift evolution at all in $\eta^{\rm obs}$, confirming at a very high level of
confidence that $b$ should be zero, as expected from the CDD. We
should also point out that this test of the CDD spans an unusually large
redshift range, all the way out to $z\sim 2.5$, which was not possible
using solely Type Ia SN and cluster data.

\begin{figure}
\centerline{
\includegraphics[angle=0,scale=0.6]{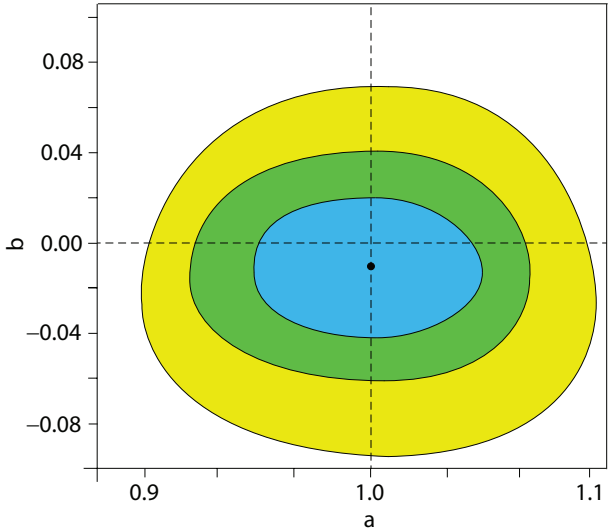}}
\vskip 0.2in
\caption{One (blue), two (green), and three (yellow) $\sigma$ confidence regions
associated with the optimized parameter $b=-0.01$ and chosen value $a=1$ in 
Equation~(9).}
\end{figure}

\section{Using the CDD to compare $R_{\rm h}=ct$ with $\Lambda$CDM}
Let us now consider a simple application of this result to model 
selection and use the CDD data to test the $R_{\rm h}=ct$ and 
$\Lambda$CDM cosmologies. Of course, the angular size measurements 
shown in fig.~1, from which the angular-diameter distances are derived, 
are independent of any particular cosmological model. 
The optimization of cosmological parameters must therefore 
be carried out by finding the best fit to the luminosity
distance that best accounts for the CDD in Equations~(9) 
and (11). This presents us with a choice of either simultaneously
fitting the HIIGx and GEHR Hubble diagram and the CDD,
or making the model selection less dependent on data (such as HIIGx and GEHR)
and simply identifying the parameters that best account for
Equations~(9) and (11). We should point out that our previous analysis
of the HIIGx and GEHR Hubble diagram (Wei et al. 2016) already demonstrated
a strong preference for $R_{\rm h}=ct$ over $\Lambda$CDM, so were we to 
include both the HII and CDD data in our fits, we would see an 
{\it a priori} bias towards $R_{\rm h}=ct$, in spite of the impact 
introduced by the CDD relation. For this paper, we therefore take a
streamlined approach and simply optimize the model parameters based
on the CDD data shown in fig.~3 on their own.

For the standard model, we assume a spatially flat cosmology with cosmological constant 
$\Lambda$. As such, the luminosity distance in $\Lambda$CDM is given as
\begin{equation}
d_{L}^{\Lambda {\rm CDM}}(z) = {c\over H_{0}}(1+z)
\int_{0}^{z}{dz\over\sqrt{\Omega_{\rm m}(1+z)^{3}+\Omega_\Lambda}}\;,
\end{equation}
in which we have assumed a negligible radiation energy density in the local Universe. 
Also, $\Omega_{\rm m}\equiv \rho_{\rm m}/\rho_c$ is today's matter density in terms
of the critical density $\rho_c\equiv 3c^2H_0^2/8\pi G$ and Hubble constant $H_0$,
and $\Omega_\Lambda=1-\Omega_{\rm m}$. The luminosity distance in the
$R_{\rm h}=ct$ universe (Melia 2003, 2007, 2013a, 2016a, 2017a; Melia \& Abdelqader 2009;
Melia \& Shevchuk 2012) is given by the much simpler expression
\begin{equation}
d_L^{R_{\rm h}=ct}(z)={c\over H_0}(1+z)\ln(1+z)\;.
\end{equation}

\begin{table}
\begin{center}
{\footnotesize
\caption{Model Selection based on the CDD}
\begin{tabular}{lcccc}
&&&& \\
\hline\hline
&&&& \\
Model& $\Omega_{\rm m}$ & $\eta_0$ & BIC & Likelihood  \\
&&&& \\
\hline
&&&& \\
$R_{\rm h}=ct$            &---& $0.51\pm0.03$ & $3.0$  & $82.3\%$ \\
&&&& \\
$\Lambda$CDM              & $0.30^{+0.26}_{-0.07}$ & $0.52\pm 0.03$ & $6.0$ & $17.7\%$ \\
&&&& \\
\hline\hline
\end{tabular}
}
\end{center}
\end{table}

As was the case in Equation~(8), we do not yet have an absolute measure
of the core size ${\ell}_{\rm core}$. In writing an expression for $\eta^{\rm th}(z)$,
however, one cannot optimize ${\ell}_{\rm core}$ and $H_0$ separately, and
we therefore put 
\begin{equation}
\eta^{\rm th}(z)=\eta_0\;{c\,\theta_{\rm core}^{-1}(z)\over H_0\,d^{\rm th}_L(z)}(1+z)^2\;,
\end{equation}
where $\eta^{\rm th}(z)$ represents the CDD for either $\Lambda$CDM
or $R_{\rm h}=ct$, in terms of the corresponding luminosity distance $d_L^{\Lambda{\rm CDM}}(z)$
or $d_L^{R_{\rm h}=ct}(z)$, and the quantity $\eta_0$ in this expression 
subsumes the constants ${\ell}_{\rm core}$, $H_0$ and $c$. 
Finding a best fit for $\Lambda$CDM therefore means optimizing the free parameters 
$\Omega_{\rm m}$ and $\eta_0$ to minimize the $\chi^2$ function 
\begin{equation}
\chi^2\equiv \sum_{i=1}^{20} {\left(a+bz_i-\eta^{\rm th}(z_i)\right)^2\over
\sigma_\eta(z_i)^2}\;,
\end{equation}
where $z_i$ are the individual redshift values in fig.~3, and $\sigma_\eta(z_i)$ is
the GP standard deviation calculated in Equation~(10). The analogous situation with
$R_{\rm h}=ct$ requires an optimization of the sole parameter $\eta_0$.

\begin{table*}
  \caption{Model Comparisons between $R_{\rm h}=ct$ and $\Lambda$CDM}
  \centering
  \begin{tabular}{lll}
&& \\
    \hline
\hline
&& \\
Test or Observational Conflict/Tension& Outcome & Reference\\
&& \\
\hline
&& \\
Angular correlation function of the CMB& $R_{\rm h}=ct$ fits it very well; standard inflationary $\Lambda$CDM misses by $>>3 \sigma$& Melia \& L\'opez-Corredoira (2018)\\
Massive halo growth at $4\lesssim z\lesssim 10$ &Data consistent with $R_{\rm h}=ct$; $\Lambda$CDM misses by $\sim 10^4$&
Steinhardt et al. (2016)\\
&&Yennapureddy \& Melia (2018b)\\
Electroweak Horizon Problem & $R_{\rm h}=ct$ does not have it; $\Lambda$CDM has no solution&Melia (2018b) \\
Missing progenitors of high-$z$ quasars&In tension with $\Lambda$CDM, but consistent
with timeline in $R_{\rm h}=ct$&Fatuzzo \& Melia (2017)\\
Angular-diameter distance test with quasar cores& $R_{\rm h}=ct$ is favoured over $\Lambda$CDM with BIC
likelihood $81\%$ vs $19\%$&Melia (2018a);\\
&&Melia \& Yennapureddy (2018)\\
HII Hubble diagram&$R_{\rm h}=ct$ is favoured over $\Lambda$CDM with  BIC likelihood $93\%$
vs. $7\%$& Wei et al. (2016)\\
&&Leaf \& Melia (2018b)\\
Alcock-Paczy\'nski test with the BAO scale & $R_{\rm h}=ct$ is favoured over $\Lambda$CDM at a $2.6\sigma$ c.l.&
Melia \& L\'opez-Corredoira (2017) \\
FSRQ $\gamma$-ray luminosity function& $R_{\rm h}=ct$ is very strongly favoured over $\Lambda$CDM with $\Delta\gg10$ &Zeng et al. (2016) \\
QSO Hubble diagram $+$ Alcock-Paczy\'nski& $R_{\rm h}=ct$ is about 4 times more likely than $\Lambda$CDM to be correct&
L\'opez-Corredoira et al. (2016) \\
Constancy of the cluster gas mass fraction & $R_{\rm h}=ct$ is favoured over $\Lambda$CDM with BIC likelihood $95\%$ vs $5\%$&Melia (2016b) \\
Cosmic Chronometers&$R_{\rm h}=ct$ is favoured over $\Lambda$CDM with BIC likelihood $95\%$ vs $5\%$&Melia \& Maier (2013); \\
&&Melia \& McClintock (2015a)\\
Cosmic age of old clusters&$\Lambda$CDM can't accommodate high-$z$ clusters, but $R_{\rm h}=ct$ can&Yu \& Wang (2014) \\
High-$z$ quasars&Evolution timeline fits within $R_{\rm h}=ct$, but not $\Lambda$CDM&Melia (2013b,2018c); \\
&&Melia \& McClintock (2015b) \\
The AGN Hubble diagram& $R_{\rm h}=ct$ is favoured over $\Lambda$CDM with BIC likelihood $96\%$ vs $4\%$&Melia (2015) \\
Age vs. redshift of old passive galaxies& $R_{\rm h}=ct$ favoured over $\Lambda$CDM with BIC likelihood $80\%$ vs $20\%$&Wei et al. (2015a) \\
Type Ic superluminous supernovae & $R_{\rm h}=ct$ is favoured over $\Lambda$CDM with BIC likelihood $80\%$ vs $20\%$&Wei et al. (2015b) \\
The SNLS Type Ia SNe&$R_{\rm h}=ct$ is favoured over $\Lambda$CDM with BIC likelihood $90\%$ vs $10\%$&Wei et al. (2015c) \\
Angular size of galaxy clusters&$R_{\rm h}=ct$ is favoured over $\Lambda$CDM with BIC likelihood $86\%$ vs $14\%$&Wei et al. (2015d) \\
Strong gravitational lensing galaxies&Both models fit the data very well due to the bulge-halo `conspiracy'& Melia et al. (2015) \\
&&Leaf \& Melia (2018a)\\
Time delay lenses&$R_{\rm h}=ct$ is favoured over $\Lambda$CDM with BIC likelihood $80\%$ vs $20\%$&Wei et al. (2014a) \\
High-$z$ galaxies&Evolution timeline fits within $R_{\rm h}=ct$, but not $\Lambda$CDM&Melia (2014a) \\
GRBs $+$ star formation rate&$R_{\rm h}=ct$ is favoured over $\Lambda$CDM with AIC likelihood $70\%$ vs $30\%$&Wei et al. (2014b) \\
High-$z$ quasar Hubble diagram&$R_{\rm h}=ct$ is favoured over $\Lambda$CDM with BIC likelihood $85\%$ vs $15\%$&Melia (2014b) \\
GRB Hubble diagram&$R_{\rm h}=ct$ is favoured over $\Lambda$CDM with BIC likelihood $96\%$ vs $4\%$&Wei et al. (2013) \\
&& \\
\hline\hline
  \end{tabular}
\end{table*}

A selection tool commonly used to differentiate between competing models (see,
e.g., Melia \& Maier 2013, and references cited therein) is the Bayes Information
Criterion, ${\rm BIC}\equiv\chi^2+k(\ln n)$, where $n$ is the number of data points and
$k$ is the number of free parameters (Schwarz 1978). For comprehensive model 
selection, one would probably use the full Bayesian evidence, rather than the BIC
approximation, but this is not really necessary here, since we are merely providing 
a demonstration of how $\eta(z)$ may be used for this purpose. This is why we are 
allowing $\Lambda$CDM to have only one free parameter ($\Omega_{\rm m}$), and 
assuming flatness and a cosmological constant. Given the robustness of the results 
shown in Table~1 below, it is very unlikely that the percentage likelihoods would be 
reversed, or even changed significantly, with the more in-depth analysis. When 
$n\gg k$, as we have here, the BIC approximates the computation of the (logarithm of the) 
`Bayes factor' for deciding between models (Schwarz 1978; Kass \& Raftery 1995). In this 
limit, the posterior distribution typically becomes increasingly peaked, 
and Gaussian in shape. As long as one may assume that the parameters have a distribution
that is unimodal and roughly Gaussian, the Bayes factor between two competing models 
can be calculated to high accuracy from the quotient of their respective (maximized) 
likelihoods. Using Laplace's method to approximate definite integrals of increasingly
peaked integrands, the Kass \& Raftery argument is similar to Stirling's approach of 
calculating an asymptotic approximation to $n!$ when $n>>1$, given by his famous 
formula.

With ${\rm BIC}_\alpha$
characterizing model $\mathcal{M}_\alpha$, the unnormalized confidence that this 
model is true is the Bayes weight $\exp(-{\rm BIC}_\alpha/2)$. 
Thus, model $\mathcal{M}_\alpha$ has likelihood 
\begin{equation}
P(\mathcal{M}_\alpha)= \frac{\exp(-{\rm BIC}_\alpha/2)}
{\exp(-{\rm BIC}_1/2)+\exp(-{\rm BIC}_2/2)}
\end{equation}
of being the correct choice when dealing with a one-on-one comparison. Another
way to think of this is in terms of the difference 
$\Delta \rm BIC \equiv {\rm BIC}_2\nobreak-{\rm BIC}_1$, which represents the 
extent to which $\mathcal{M}_1$ is favored over~$\mathcal{M}_2$. The outcome 
$\Delta\equiv$ BIC$_1-$ BIC$_2$ is judged `positive' in the range $\Delta=2-6$, 
`strong' for $\Delta=6-10$, and `very strong' for $\Delta>10$.

Our model comparison is summarized in Table~1, which displays several promising
features. First, the optimization of $\eta_0$ appears to be essentially independent
of the model, which suggests that both $R_{\rm h}=ct$ and $\Lambda$CDM provide
adequate fits to the CDD in Equations~(9) and (11). Second,
the optimized matter density $\Omega_{\rm m}=0.30^{+0.26}_{-0.07}$ in $\Lambda$CDM 
is remarkably consistent with the value $\Omega_{\rm m}=0.308\pm 0.012$ measured 
by {\it Planck} (Planck Collaboration 2016). All of this represents an internal 
self-consistency that reinforces the validity of the CDD in
Equations~(1), (9) and (11), particularly with regard to our approach in this paper of using the 
quasar compact cores to measure $d_A(z)$ and the reconstruction of the HIIGx 
and GEHR Hubble diagram using Gaussian Processes to measure $d_L(z)$. 

Nonetheless, a notable difference does emerge between these two models, directly 
attributable to the number of free parameters $k$. Once $H_0$ is subsumed into $\eta_0$, 
$R_{\rm h}=ct$ has no additional degrees of freedom to use in fitting the $\eta^{\rm obs}(z)$ 
data in fig.~3. This is quite constraining compared to $\Lambda$CDM, in which one 
may adjust $\Omega_{\rm m}$ to improve the fit. This added flexibility is reflected
in the standard model's larger BIC, a consequence of the greater penalty imposed
by the information criterion on the less parsimonious models. The magnitude of 
the difference $\Delta \rm BIC=3.0$ indicates that the evidence in
favour of $R_{\rm h}=ct$ is positive. As a result, the likelihood of $R_{\rm h}=ct$ 
being the correct cosmology, based on the CDD relation, is $\sim 82.3\%$
compared with only $\sim 17.7\%$ for $\Lambda$CDM. 

Given this outcome, it may be helpful to compare this prioritization with
the results of other comparative tests that have been reported in the literature
over the past decade (see Table~2). As one may see from this list, the fact that
the CDD relation tends to favour $R_{\rm h}=ct$ over $\Lambda$CDM affirms the
general trend seen earlier with measurements taken at both low and high redshifts,
using a broad range of sources and signatures, including integrated distances
and times, and also the differential expansion rate. Perhaps the most notable
example of this comparison has to do with the temperature and electroweak
horizon problems that require fixes to make $\Lambda$CDM work properly in the
early Universe. Inflation may solve the former, but there is currently no 
established resolution of the latter (Melia 2018b). On the other hand, 
$R_{\rm h}=ct$ has neither, because it avoids the early deceleration present
in the standard model that produces these excessively small horizons in the
first place. 

Of course, there is still much to be done before one can claim
that $R_{\rm h}=ct$ is the correct model instead of $\Lambda$CDM. In this
picture, dark energy is dynamic, not a cosmological constant, so new physics
beyond the standard model of particle physics is required. It will also be
essential to understand how fluctuations are produced in this cosmology,
and whether they grow to properly account for the large-scale structure
we see today. These are just a few of the many remaining issues that must
be resolved going forward. The test reported in this paper helps to remove
at least some of the uncertainty with the underlying physics, in this case
having to do with the distance duality relation, which continues to build
the evidence in favour of one model over the other.

\section{Conclusion}
The approach we have introduced in this paper to test the CDD appears 
to be an improvement over previous methods for several reasons. Once a suitable
sample of compact quasar cores is selected, the use of these sources as standard
rulers is rather clean---not fraught with issues, such as a complex interior structure
in galaxy clusters that produces irreducible scatter in the measurement of an
angular diameter distance. For example, consider the disparity produced by the
comparison of two different cluster samples in Holanda et al. (2010, 2012), who
reached widely different conclusions regarding the validity of the CDD, depending
on whose data one chooses to use for the reciprocity relation.

Second, the Hubble diagram based on HIIGx and GEHR measurements may be used to
determine the luminosity distance without the assumption of any 
particular cosmological model. This
application is made possible through the use of Gaussian Processes to 
reconstruct the distance modulus as a function of redshift. Third, these
two sets of data allow us to measure the CDD over a significantly larger 
redshift range than was possible with Type Ia SNe. Not only have we
confirmed the CDD with a higher precision than before, but we have done so
all the way out to $z\sim 2.5$. 

Finally, we have demonstrated the practicality of this outcome by using it
to test two competing cosmologies. We have shown that the measured CDD data
favour the $R_{\rm h}=ct$ universe over $\Lambda$CDM with a likelihood
of $\sim 82.3\%$ versus only $\sim 17.7\%$, confirming the results of
previously published model comparisons based on over 23 other kinds of 
data, which are summarized in Table~1 of Melia (2017b). In this regard, 
it is worth comparing this result with that of another recent application 
of the CDD by Hu \& Wang (2018) to test the $R_{\rm h}=ct$ and $\Lambda$CDM 
cosmologies. These workers based their analysis on an entirely different 
approach, with a significantly different conclusion. 

How does one reconcile such different outcomes? In our case, the CDD was
first confirmed independently of any cosmological model, and 
the results summarized in Equation~(11) and fig.~4 are entirely consistent 
with $\eta(z)=1$. In using the CDD to test our models, there was no need to 
include model fits to the compact quasar core and HIIGx GEHR data themselves, 
which would have obscured the true outcome based on the CDD itself. As such, 
there were no complications arising from the effects on $d_L(z)$ from the merger 
of sub-samples of sources with unknown intrinsic dispersions, or with non-uniform
calibrations. Our comparison is clean and is directly based on the CDD itself.

In contrast, Hu \& Wang (2018) did not use a cosmology-independent 
approach to examine the CDD, opting instead to base their analysis on older methods
using galaxy clusters and Type Ia SNe. The outcome of their comparison
is therefore heavily biased by the impact of fits to the cluster and SN
data themselves, rather than being a true reflection of the CDD. Unfortunately,
these fits---particularly to the Type Ia SN data---are heavily tainted by
the many problems encountered with model comparisons based on such observations,
as described in several published accounts (see, e.g., Kim 2011; Wei et al.
2015). Of particular concern with their work is the fact that their results
are strongly dependent on the Hubble constant, which they demonstrated by
comparing two values, though failing to optimize $H_0$ separately for each
model. The most serious drawback with their approach, however, is simply
ignoring the unknown systematic differences between sub-samples merged to
produce the overall SN catalog. As demonstrated in Wei et al. (2015),
one should ideally use a single SN sample, with a homogeneous calibration
and systematics for all the data.  But even then, that method of
measuring the CDD is inferior to a true cosmology-independent 
approach, as we have in this paper, which produces an unbiased determination 
of $\eta(z)$ for model comparisons.

Finally, with an eye to possible future applications of this
work, we recall several cautionary remarks we have made concerning the
dependence of this work on possible unknown systematics in the quasar
core and HII galaxy data. Given how well the test of the CDD has 
turned out with these sources, perhaps one should turn this procedure
around and use the CDD to evaluate the internal self-consistency of
the astrophysical models used for the radio emission and compact structure 
of the former, and the HII line emission and velocity dispersion in the
latter. At the very least, an application of the CDD to such sources
may delimit the extent to which any unaccounted for systematics and 
unknowns are impacting their observed spectra and luminosities. 

In conclusion, the fact that the CDD is confirmed by the observations
hardly surprises anyone. After all, many cosmological measurements tacitly
assume its validity anyway. Nonetheless, the cosmology-independent 
approach we have used in this paper has provided a compelling demonstration 
that distance duality is indeed realized in nature.

\section*{Acknowledgments} I am grateful to the anonymous
referee for suggesting several improvements to this manuscript. I am also 
grateful to Amherst College 
for its support through a John Woodruff Simpson Lectureship, and to Purple 
Mountain Observatory in Nanjing, China, for its hospitality while part of 
this work was being carried out. This work was partially supported by grant 
2012T1J0011 from The Chinese Academy of Sciences Visiting Professorships for 
Senior International Scientists, and grant GDJ20120491013 from the Chinese 
State Administration of Foreign Experts Affairs.

\label{lastpage}


\vfill\newpage

\begin{thebibliography}{99}

\bibitem{Adler1971} Adler, S. L., 1971, AnPhy, 67, 599
\bibitem{Amanullah2010} Amanullah, R., Lidman, C., Rubin, D. et al. 2010, ApJ, 716, 712
\bibitem{Bassett2004a} Bassett, B. A. \& Kunz, M., 2004a, ApJ, 607, 661
\bibitem{Bassett2004b} Bassett, B. A. \& Kunz, M., 2004b, PRD, 69, 101305
\bibitem{Belanger2004} B\'elanger, G., Goldwurm, A., Goldoni, P., Paul, J., Terrier, R., Falanga, M. et al.,
2004, ApJL, 601, L163
\bibitem{Bergeron1977} Bergeron, J., 1977, ApJ, 211, 62
\bibitem{Bernardis2006} Bernardis, F. D., Giusarma, E. \& Melchiorri, A., 2006, IJMP-D, 15, 759
\bibitem{Blanchard06} Blanchard, A., 2006, in: Current issues in Cosmology, 
eds. J.-C. Pecker and J. V. Narlikar, (Cambridge University Press, Cambridge U.K., p.~76
\bibitem{Blandford1979} Blandford, R. D. \& K\"onigl, A., 1979, ApJ, 232, 34
\bibitem{Bordalo2011} Bordalo, V. \& Telles, E., 2011, ApJ, 735, 52
\bibitem{Bosch2002} Bosch, G., Terlevich, E. \& Terlevich, R., 2002, MNRAS, 329, 481
\bibitem{Burrage2008} Burrage, C., 2008, PRD, 77, 043009
\bibitem{Cao2017} Cao, S. et al., 2017, JCAP, 02, id. 012
\bibitem{Chan2009} Chan, C.-K., Liu, S., Fryer, C. L., Psaltis, D., \"Ozel, F., Rockefeller, G. \& Melia, F.,
2009, ApJ, 701, 521
\bibitem{Chashchina15} Chashchina, O. I. and Silagadze, Z. K., 2015, Universe, 1, 307
\bibitem{Chavez2012} Ch{\'a}vez, R., Terlevich, E., Terlevich, R., Plionis, M., 
Bresolin, F., Basilakos, S. \& Melnick, J., 2012, MNRAS Lett, 425, L56
\bibitem{Chavez2014} Ch{\'a}vez, R., Terlevich, R., Terlevich, E., Bresolin, F., 
Melnick, J., Plionis, M. \& Basilakos, S., 2014, MNRAS, 442, 3565
\bibitem{Chavez2016} Ch\'avez, R., Plionis, M., Basilakos, S., 
et~al., 2016, MNRAS, 462, 2431
\bibitem{Chen1995} Chen, P., 1995, PRL, 74, 634
\bibitem{Crocker2011} Crocker, R. M., Jones, D. I., Aharonian, F., Law, C. J., Melia, F. \& Ott, J.,
2011, MNRAS Letters, 411, L11
\bibitem{Deffayet2000} Deffayet, C. \& Uzan, J.-P., 2000, PRD, 62, 063507
\bibitem{Ellis1971} Ellis, G.F.R., 1971, in Proc. School "Enrico Fermi", Ed. R. K.  
Sachs, New York
\bibitem{Ellis2013} Ellis, G.F.R., Poltis, R., Uzan, J.-P. \& Weltman, A., 2013, PRD, 87, 103530
\bibitem{Etherington1933} Etherington, I.M.H., 1933, Philos. Mag., 15, 761
\bibitem{Fatuzzo2017} Fatuzzo, M. \& Melia, F., 2017, ApJ, 846, 129
\bibitem{Fuentes2000} Fuentes-Masip, O., Mu{\~n}oz-Tu{\~n}{\'o}n, C., 
Casta{\~n}eda, H.~O. \& Tenorio-Tagle, G., 2000, AJ, 120, 752
\bibitem{Gurvits1994} Gurvits, L. I., 1994, ApJ, 425, 442
\bibitem{Gurvits1999} Gurvits, L. I., Kellermann, K. I. \& Frey, S., 1999, A\&A, 342, 378
\bibitem{Holanda2010} Holanda, R.F.L., Lima, J. A. \& Ribeiro, M. B., 2010, ApJL, 722, L233
\bibitem{Holanda2012} Holanda, R.F.L., Lima, J. A. \& Ribeiro, M. B., 2012, A\&A, 538, A131 
\bibitem{Hu2018} Hu, J. \& Wang, F.-Y., 2018, MNRAS, in press (arXiv:1804.06606)
\bibitem{Jackson2004} Jackson, J. C., 2004, JCAP, 11, id. 007
\bibitem{Jackson2008} Jackson, J. C., 2008, MNRAS, 390, L1
\bibitem{Jackson1997} Jackson, J. C. \& Dodgson, M., 1997, MNRAS, 285, 806
\bibitem{JacksonJannetta2006} Jackson, J. C. \& Jannetta, A. L.,2006, JCAP, 11, 002
\bibitem{Kass1995} Kass, R. E. \& Raftery, A. E., 1995, J. Amer. Stat. Ass., 90, 773
\bibitem{Kellermann1993} Kellermann K.~I., 1993, Nature, 361, 134
\bibitem{Kim2011} Kim, A. G., 2011, PASP, 123, 230
\bibitem{Khedekar2011} Khedekar, S. \& Chakraborti, S., 2011, PRL, 106, 221301
\bibitem{Khoury2004} Khoury, J. \& Weltman, A., 2004, PRL, 93, 171104
\bibitem{Kunth2000} Kunth, D. \& {\"O}stlin, G., 2000, A\&ARv, 10, 1
\bibitem{LaViolette12} LaViolette, P. A., 2012, Subquantum kinetics: The Alchemy of
Creation, 4th ed. (Starlane Pub., Niskayana NY)
\bibitem{LeafMelia2018a} Leaf, K. \& Melia, F., 2018a, MNRAS, 478, 5104
\bibitem{LeafMelia2018b} Leaf, K. \& Melia, F., 2018b, MNRAS, 474, 4507
\bibitem{Li2011} Li, Z., Wu, P. \& Yu, H., 2011, ApJL, 729, L14
\bibitem{Liao2015} Liao, K., Avgoustidis, A. \& Li, Z., 2015, PRD, 92, 123539
\bibitem{Liao2016} Liao, K., Li, Z., Cao, S., Biesiada, M., Zheng, X. \& Zhu, Z.-H., 2016, ApJ, 822, 74
\bibitem{Liu2001} Liu, S. \& Melia, F., 2001, ApJL, 561, L77
\bibitem{Lopez2016} L\'opez-Corredoira, M., Melia, F., Lusso, E. and Risaliti, G., 2016, IJMP-D, 25, id. 1650060
\bibitem{Ma2016} Ma, C. \& Corasaniti, P.-S., 2016, submitted (arXiv:1604.04631)
\bibitem{Mania2012} Mania, D. \& Ratra, B., 2012, PhLB, 715, 9
\bibitem{Melia1992} Melia, F., Jokipii, J. R. \& Narayanan, A., 1992, ApJL, 395, L87
\bibitem{Melia2003} Melia F., 2003, ``The Edge of Infinity: Supermassive Black Holes 
in the Universe" (New York: Cambridge University Press), 158--171
\bibitem{Melia2007} Melia, F., 2007, MNRAS, 382, 1917
\bibitem{Melia2013a} Melia, F., 2013a, A\&A, 553, A76
\bibitem{Melia2013b} Melia, F., 2013b, ApJ, 764, 72
\bibitem{Melia2014a} Melia, F., 2014a, AJ, 147, 120
\bibitem{Melia2014b} Melia, F., 2014b, JCAP, 01, 027
\bibitem{Melia2015} Melia, F., 2015, ASpSci 359, 34
\bibitem{Melia2016a} Melia, F., 2016a, Front. Phys., 11, 119801
\bibitem{Melia2016b} Melia, F., 2016b, Proc. R. Soc. A,  472, 20150765
\bibitem{Melia2017a} Melia, F., 2017a, Front. Phys., 12, 129802
\bibitem{Melia2017b} Melia, F., 2017b, MNRAS, 464, 1966
\bibitem{Melia2018a} Melia, F., 2018a, EPL, 123, id. 39001  
\bibitem{Melia2018b} Melia, F., 2018b, EPJ-C, in press (arXiv:1809.02885)
\bibitem{Melia2018c} Melia, F., 2018c, A\&A, 615, A113
\bibitem{MeliaAbdelqader2009} Melia, F. \& Abdelqader, M., 2009, IJMP-D, 18, 1889
\bibitem{Melia1989} Melia, F. \& K\"onigl, A., 1989, ApJ, 340, 162
\bibitem{MeliaLopez2016} Melia, F. \& L\'opez-Corredoira, M., 2017, IJMP-D, 26, id. 1750055
\bibitem{MeliaLopez2018} Melia, F. \& L\'opez-Corredoira, M., 2018, A\&A, 610, A87
\bibitem{MeliaMaier2013} Melia F. and Maier R.~S., 2013, MNRAS, 432, 2669
\bibitem{MeliaMcClintock2015a} Melia, F. and McClintock, T. M., 2015a, AJ, 150, id. 119
\bibitem{MeliaMcClintock2015b} Melia, F. and McClintock, T. M., 2015b, Proc. R. Soc. A,
\bibitem{MeliaShevchuk2012} Melia, F. \& Shevchuk, A.,  2012, MNRAS, 419, 2579
\bibitem{MeliaWei2015} Melia, F., Wei, J.-J. \& Wu, X., 2015, AJ, 149, 2
\bibitem{MeliaYennapureddy2018} Melia, F. \& Yennapureddy, M. K., 2018, MNRAS, 480, 2144
\bibitem{Melnick1987} Melnick, J., Moles, M., Terlevich, R. \& Garcia-Pelayo, J.-M., 1987, MNRAS, 226, 849
\bibitem{Melnick1988} Melnick, J., Terlevich, R. \& Moles, M., 1988, MNRAS, 235, 297
\bibitem{Melnick2000} Melnick, J., Terlevich, R. \& Terlevich, E., 2000, MNRAS, 311, 629
\bibitem{Meng2012} Meng, X.-L., Zhang, T.-J., Zhan, H. \& Wang, X., 2012, ApJ, 745, 98
\bibitem{Nair2011} Nair, R., Jhingan, S. \& Jain, D., 2011, JCAP, 05, 023
\bibitem{Nayakshin1998} Nayakshin, S. \& Melia, F., 1998, ApJS, 114, 269
\bibitem{Planck2016} Planck Collaboration et al., 2016, A\&A, 594, id A13
\bibitem{Plionis2011} Plionis, M., Terlevich, R., Basilakos, S., Bresolin, F., 
Terlevich, E., Melnick, J. \& Chavez, R., 2011, MNRAS, 416, 2981
\bibitem{Preston1985} Preston, R. A., 1985, AJ, 90, 1599
\bibitem{Raffelt1999} Raffelt, G. G., 1999, ARNPS, 49, 163
\bibitem{Santos2008} Santos, R. C. \& Lima, J.A.S., 2008, PRD, 77, 083505
\bibitem{Schwarz1978} Schwarz, G.\ 1978, Ann. Statist., 6, 461
\bibitem{Searle1972} Searle, L. \& Sargent, W.L.W., 1972, ApJ, 173, 25
\bibitem{Seikel2012} Seikel, M., Clarkson, C. \& Smith, M., 2012, JCAP, 06, 036S
\bibitem{Siegel2005} Siegel, E.~R., Guzm{\'a}n, R., Gallego, J.~P., 
Ordu{\~n}a-L{\'o}pez, M. \& Rodr{\'{\i}}guez-Hidalgo, P., 2005, MNRAS, 356, 1117
\bibitem{sikivie1983} Sikivie, P., 1983, PRL, 51, 1415
\bibitem{Steinhardt2016} Steinhardt, C. L.,  Capak, P., Masters, D. \& Speagle, J. S., 2016, ApJ, 824, 1
\bibitem{Telles2003} Telles, E., 2003, ASPC, 297, 143
\bibitem{Terlevich1981} Terlevich, R. \& Melnick, J., 1981, MNRAS, 195, 839
\bibitem{Terlevich2015} Terlevich, R., Terlevich, E., Melnick, J., Ch{\'a}vez, R., 
Plionis, M., Bresolin, F. \& Basilakos, S., 2015, MNRAS, 451, 3001
\bibitem{Trap2011} Trap, G., Goldwurm, A., Dodds-Eden, K., Weiss, A., Terrier, R., Ponti, G. et al.,
2011, A\&A, 528, id. A140
\bibitem{Uzan2004} Uzan, J.-P., Aghanim, N. \& Mellier, Y., 2004, PRD, 70, 083533
\bibitem{Vauclair03} Vauclair, S. C. et al., 2003, A\&A, 412, L37
\bibitem{Vishwakarma2001} Vishwakarma, R. G., 2001, CQG, 18, 1159
\bibitem{Vishwakarma13} Vishwakarma, R. G., 2013, Phys. Scr., 87, 055901
\bibitem{Wei2013} Wei, J.-J., Wu, X. and Melia, F., 2013, ApJ, 772, id. 43
\bibitem{Wei2014a} Wei, J.-J., Wu, X. and Melia, F., 2014a, ApJ, 788, id. 190
\bibitem{Wei2014b} Wei, J.-J., Wu, X., Melia, F., Wei, D.-M. and  Feng, L.-L., 2014b, MNRAS, 439, 3329
\bibitem{Wei2015a} Wei, J.-J., Wu, X., Melia, F., Wang, F.-Y. and Yu, H., 2015a, AJ, 150, id. 35
\bibitem{Wei2015b} Wei, J.-J., Wu, X. and Melia, F., 2015b, AJ, 149, id. 165
\bibitem{Wei2015c} Wei, J.-J., Wu, X., Melia, F. and Maier, R. S., 2015c, AJ, 149, id. 102
\bibitem{Wei2015d} Wei, J.-J., Wu, X. and Melia, F., 2015d, MNRAS, 447, 479
\bibitem{Wei2016} Wei, J.-J., Wu, X.-F. \& Melia, F., 2016, MNRAS, 463, 1144
\bibitem{Yang2013} Yang, X., Yu, H.-R., Zhang, Z.-S. \& Zhang, T.-J., 2013, ApJL, 777, L24
\bibitem{Yennapureddy2017} Yennapureddy, M. K. \& Melia, F., 2017, JCAP, 11, 029
\bibitem{Yennapureddy2018a} Yennapureddy, M. K. \& Melia, F., 2018a, EPJ-C, 78, 258
\bibitem{Yennapureddy2018b} Yennapureddy, M. K. \& Melia, F., 2018b, PDU, 20, 65 
\bibitem{Yu2014} Yu, H. \& Wang, F. Y., 2014, Eur. Phys. J. C, 74, 3090
\bibitem{Zeng2016} Zeng, H., Melia, F. \& Zhang, L., 2016, MNRAS, 462, 3094

\end{thebibliography}
\end{document}